\documentclass[a4paper,11pt]{article}

\usepackage{jinstpub} 

\usepackage{lineno}

\usepackage{placeins}

\usepackage{comment}

\title{Cryogenic front-end amplifier design for large SiPM arrays in the DUNE FD1-HD photon detection system}

\author[a,b]{C.~Brizzolari,}
\author[a,b]{P.Carniti,}
\author[a]{C.~Cattadori,}
\author[a,b]{E.~Cristaldo,}
\author[c]{A.~de la Torre Rojo,}
\author[a,b]{M.~Delgado,}
\author[a,b]{A.~Falcone,}
\author[d]{K.~Francis,}
\author[e,f]{N.~Gallice,}
\author[a]{C.~Gotti,}\note{Corresponding author.}
\author[g,h]{M.~Guarise,}
\author[i,l]{E.~Montagna,}
\author[i,l]{A.~Montanari,}
\author[a]{G.~Pessina,}
\author[i,l]{M.~Pozzato,}
\author[m]{J.~Smolik,}
\author[a,b]{F.~Terranova,}
\author[g,h]{L.~Tomassetti,}
\author[c]{A.~Verdugo de Osa,}
\author[o]{D.~Warner,}
\author[n]{J.~Zalesak,}
\author[e]{A.~Zani,}
\author[n]{J.~Zuklin,}
\author[d]{and V.~Zutshi}

\affiliation[a]{Istituto Nazionale di Fisica Nucleare, Sezione di Milano Bicocca, I-20126 Milano, Italy}
\affiliation[b]{Universit{\`a} di Milano-Bicocca, I-20126 Milano, Italy}
\affiliation[c]{CIEMAT, Centro de Investigaciones Energ{\'e}ticas, Medioambientales y Tecnol{\'o}gicas, E-28040 Madrid, Spain}
\affiliation[d]{Northern Illinois University, DeKalb, IL 60115, USA}
\affiliation[e]{Istituto Nazionale di Fisica Nucleare, Sezione di Milano, I-20133 Milano, Italy}
\affiliation[f]{Universit{\`a} degli Studi di Milano, I-20133 Milano, Italy}
\affiliation[g]{Istituto Nazionale di Fisica Nucleare, Sezione di Ferrara, I-44122 Ferrara, Italy}
\affiliation[h]{Universit{\`a} di Ferrara, I-44122 Ferrara, Italy}
\affiliation[i]{Istituto Nazionale di Fisica Nucleare, Sezione di Bologna, I-40127 Bologna, Italy}
\affiliation[l]{Universit{\`a} di Bologna, I-40127 Bologna, Italy}
\affiliation[m]{Czech Technical University, 115 19 Prague 1, Czech Republic}
\affiliation[n]{Institute of Physics, Czech Academy of Sciences, 182 00 Prague 8, Czech Republic}
\affiliation[o]{Colorado State University, Fort Collins, CO 80523, USA}

\emailAdd{claudio.gotti@mib.infn.it}

\abstract{The photon detection system of the first far detector (FD1-HD) of the DUNE experiment will detect scintillation photons produced by particle interactions in a kiloton-scale liquid Argon time projection chamber.
The photon detectors of choice are silicon photomultipliers (SiPM), \mbox{6$\times$6 mm$^2$} each, arranged in groups of \mbox{48}, which present a significantly low impedance to the front-end electronics.
This paper details the design of a cryogenic amplifier with exceptionally low white voltage noise of 0.37 nV$\sqrt{Hz}$, based on a silicon-germanium input transistor and a BiCMOS fully differential operational amplifier.
It yields excellent single photoelectron resolution even at low overvoltage values.
The signal rise time is below 100 ns, and the dynamic range is about 2000 photoelectrons at the typical operating overvoltage.
It draws \mbox{0.7 mA} from a single \mbox{3.3 V} supply, for a power consumption of \mbox{2.4 mW} per channel.
Simplified models were developed to predict the single photolectron signal shape and the signal to noise ratio, with a good match to measured performance.}


\begin{document}
\maketitle
\flushbottom

\section{The DUNE FD1-HD photon detection system}

The Deep Underground Neutrino Experiment (DUNE) is a next generation long baseline neutrino experiment currently under construction \cite{DUNE}.
A high intensity neutrino beam will be generated at Fermilab (Illinois, USA) and detected at SURF (South Dakota, USA) after a \mbox{1300 km} flight distance.
Physics goals include the precise measurement of neutrino oscillations, the determination of the neutrino mass hierarchy, the observation of supernova events and the search for rare decays beyond the standard model.

The construction of the far detector at SURF will see the phased deployment of four kiloton-scale liquid Argon (LAr) time projection chambers.
The first far detector (FD1), named horizontal drift (HD), will have vertical anode and cathode planes to detect charge produced in LAr by the events of interest and drifted by a horizontal electric field \cite{FD1}.
The anode planes will include photon detectors, to complement the charge information with the readout of scintillation light and provide a precise time reference for events unrelated with the beam.

\begin{figure}[t]
\centering
\includegraphics[scale=0.4]{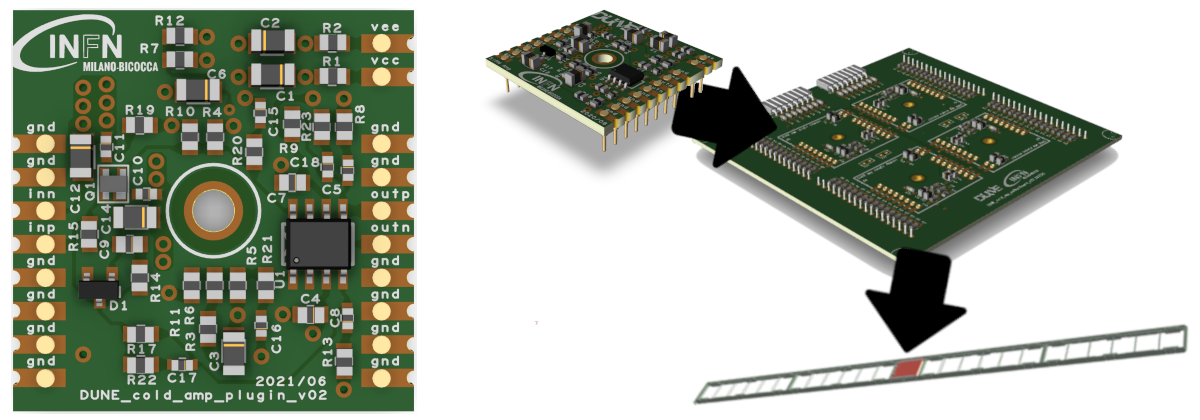}
\caption{\label{fig:ColdAmpLayout} Layout view of one amplifier, designed in a 3$\times$3 cm$^2$ printed circuit board to be plugged into the four-channel motherboard at the center of each photon detection module.}
\end{figure}

The photon detection system is arranged in photodetection modules, each made of four supercells based on the X-Arapuca technology \cite{Arapuca2, Arapuca1}.
Scintillation photons produced by particle interactions in LAr are collected, shifted to visible wavelength and guided to 48 6$\times$6 mm$^2$ silicon photomultipliers (SiPM) located along the edges of the supercells.
Each supercell constitutes a channel, read out by a low noise amplifier operating in LAr.
Figure \ref{fig:ColdAmpLayout} shows one amplifier, laid out on a 3$\times$3 cm$^2$ printed circuit board.
Four amplifiers are plugged into a motherboard, which is located at the center of the photodetector module, as shown on the right side of the same figure.
The amplified signals from four supercells in each module are driven over $\sim$30 m long transmission lines to digitizers located outside the cryostat.
This arrangement gives high flexibility in prototyping and testing, since amplifiers can be easily replaced without having to dismount the entire module.
At a later phase, after the prototypes will be fully validated, this approach might be traded for a more compact design with all four amplifiers in the same board.
A total of 6000 cold amplifier channels will be deployed in DUNE FD1-HD.

\section{SiPM electrical characteristics}

\begin{figure}[b]
\centering
\includegraphics[scale=0.75]{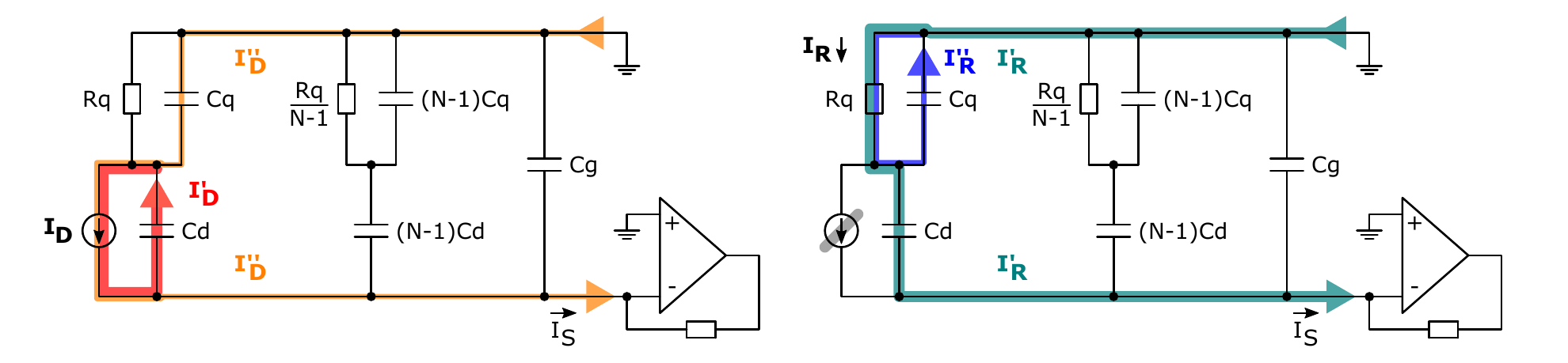}
\caption{\label{fig:EquivalentSiPM} Equivalent circuit of a SiPM read out with an amplifier with negligible input impedance. The cell discharge is shown on the left, followed by the cell recharge on the right.}
\end{figure}

From the electrical point of view, each reverse-biased cell in a SiPM can be modeled as a diode capacitance $C_d$ in series with its quenching resistor $R_q$. A stray capacitance $C_q$, whose value is a fraction of $C_d$, is in parallel with $R_q$.
Each cell also exhibits a shunt parasitic capacitance, which in parallel with the contribution of all other cells constitutes the grid capacitance $C_g$.
Figure \ref{fig:EquivalentSiPM} shows the equivalent circuit of a SiPM with $N$ cells, of which only one is fired. The remaining $N-1$ were lumped into a single branch.
The SiPM is here assumed to be read out by an ideal amplifier with negligible input impedance.
As long as this is true, the anode and cathode terminals are held at constant voltage.
The cell that generates the signal is not loaded by the other $N-1$ cells, and all the signal current goes to the amplifier.

The electrical model used here follows what is firmly established in literature for SPADs and SiPMs \cite{SiPMmodel1, SiPMmodel2, SiPMmodel3, SiPMmodel4}.
The purpose of this section is to arrive at two expressions \ref{eq:SiPMcurrent1} or \ref{eq:SiPMimpedance1} for the SiPM current and source impedance, or their simplified counterparts \ref{eq:SiPMcurrent2} or \ref{eq:SiPMimpedance2}, which will be used in section \ref{sec:Signals} to predict the expected signal shape at the output of the amplifier.

Figure \ref{fig:EquivalentSiPM} distinguishes the instantaneous discharge of the SiPM cell from its slower recovery to equilibrium conditions.
The discharge of a SiPM cell is represented by the current impulse $I_D(t)=Q \delta(t)$, where \mbox{$Q=(C_d+C_q) V_{OV}$} is the amount of charge that causes the voltage across $C_d$ to instantly drop from the operating voltage $V_{br}+V_{OV}$ to the breakdown voltage $V_{br}$, at which the avalanche is quenched.
The impulse current divides between $C_d$ and $C_q$.
The larger fraction $I_D'$ is absorbed by $C_d$, and is not seen outside the SiPM, while a smaller fraction $I_D''$ reaches the amplifier:
\begin{equation}
    I_D'(t) = \frac{C_d}{C_d+C_q} Q \delta(t),
    \hspace{1 cm}
    I_D''(t) = \frac{C_q}{C_d+C_q} Q \delta(t).
\end{equation}
After the discharge, the total capacitance of the cell $C_d+C_q$ recharges through $R_q$ with time constant $\tau_S = (C_d+C_q)R_q$.
The recharge current $I_R$ through $R_q$ is again divided according to the capacitance ratio.
The larger fraction $I_R'$ goes through $C_d$ and reaches the amplifier, while a smaller fraction $I_R''$ is absorbed by $C_q$:
\begin{equation}
    I_R'(t) = \frac{C_d}{C_d+C_q} \frac{Q e^{-t/\tau_S}}{\tau_S},
 \hspace{1 cm}
    I_R''(t) = \frac{C_q}{C_d+C_q} \frac{Q e^{-t/\tau_S}}{\tau_S}.
\end{equation}
All considered, the signal current that reaches the amplifier is
\begin{equation}
    I_S(t) = I_D''(t)+I_R'(t) =  Q \left( \frac{C_q}{C_d+C_q} \delta(t) + \frac{C_d}{C_d+C_q} \frac{e^{-t/\tau_S}}{\tau_S}   \right)
    =
    V_{OV} \left( C_q \delta(t) + C_d \frac{e^{-t/\tau_S}}{\tau_S}   \right).
    \label{eq:SiPMcurrent1}
\end{equation}
Although simplified, this model captures the main features of the SiPM signal, including its articulation in two components with the same sign: a fast spike carrying a fraction of the charge, followed by a slower component carrying the rest of the charge on a longer time scale.
In practice, the spike that is here modeled as a Dirac delta will be smoothed by stray impedances and by the finite bandwidth of the amplifier, but it might still be observable as a faster component, depending on the relative weight of $C_q$.
If $C_q$ is assumed negligible compared to $C_d$, then $\tau_S \simeq C_d R_q$ and equation \ref{eq:SiPMcurrent1} simplifies to
\begin{equation}
    I_S(t) \simeq Q \frac{e^{-t/\tau_S}}{\tau_S}   = \frac{V_{OV}}{R_q}  e^{-t/\tau_S}.
     \label{eq:SiPMcurrent2}
\end{equation}
In other words, in this case the signal current coincides with the cell recovery current $I_R$.
The limits of validity of \ref{eq:SiPMcurrent2}, which neglects $C_q$, as opposed to \ref{eq:SiPMcurrent1}, will be discussed section \ref{sec:Signals}.

In the domain of the complex frequency $s$, the source impedance of a SiPM according to the model in figure \ref{fig:EquivalentSiPM} is given by
\begin{equation}
    Z_S(s) = \frac{1+ s R_q (C_d+C_q)}{s(NC_d+C_g) + s^2 \left[R_q C_g (C_d+C_g) + N R_q C_d C_q \right]}.
     \label{eq:SiPMimpedance1}
\end{equation}
At low frequency, the source impedance is equivalent to a capacitance of value $NC_d+C_g$.
At signal frequencies, above $s \sim 1/\tau_S$, it is approximated by a resistance $R_q/N$.
At higher frequencies it is again approximated by a capacitance $N C_q+C_g$.
If $C_q$ and $C_g$ are neglected, equation \ref{eq:SiPMimpedance1} simplifies to
\begin{equation}
    Z_S(s) \simeq \frac{1+ s Rq C_d}{s N C_d}.
     \label{eq:SiPMimpedance2}
\end{equation}
Again, the limits of validity of \ref{eq:SiPMimpedance2} compared to \ref{eq:SiPMimpedance1} will be discussed section \ref{sec:Signals}.

\begin{table}[t]
\centering
\caption{\label{tab:SiPMs} Main electrical characteristics of the two SiPM models (from different vendors) used for the measurements presented in this paper.}
\smallskip
\begin{tabular}{|l|c|c|}
\hline
& Model 1 & Model 2\\
\hline
Device size & 6$\times$6 mm$^2$ & 6$\times$6 mm$^2$\\
Number of cells & 6364 & 11188\\
Cell size & 75$\times$75 $\mu$m$^2$ & 50$\times$50 $\mu$m$^2$\\
\hline
Cell characteristics at 77 K & & \\
$\cdot$ Total capacitance $C_d+C_q$ & 240 fF & 190 fF\\
$\cdot$ Quenching resistance $R_q$ & 2.2 M$\Omega$ & 3.6 M$\Omega$\\
$\cdot$ Time constant $\tau_S$ & 530 ns & 680 ns\\
\hline
Operating overvoltage & & \\
$\cdot$ 40\% PDE & +2.0 V & +3.5 V\\
$\cdot$ 45\% PDE & +2.5 V & +4.5 V\\
$\cdot$ 50\% PDE & +3.0 V & +7.0 V\\
\hline
\end{tabular}
\end{table}

Two SiPM models from different vendors have been used to validate the performance of the amplifier presented in this paper. 
Table \ref{tab:SiPMs} summarizes their main electrical characteristics.
The value of quenching resistance $R_q$ was determined from the IV curve of the SiPMs in forward bias, under the reasonable assumption that $R_q$ does not depend on the bias conditions.
The value of total capacitance $C_d + C_q$ was then extracted from the measured signal fall time $\tau_s$.
The different technology and different cell size reflect into different operating overvoltages to achieve the same photodetection efficiency (PDE).

\FloatBarrier
\section{Ganging and readout scheme}
\label{sec:Ganging}

\textbf{\begin{figure}[t]
\centering
\includegraphics[scale=0.75]{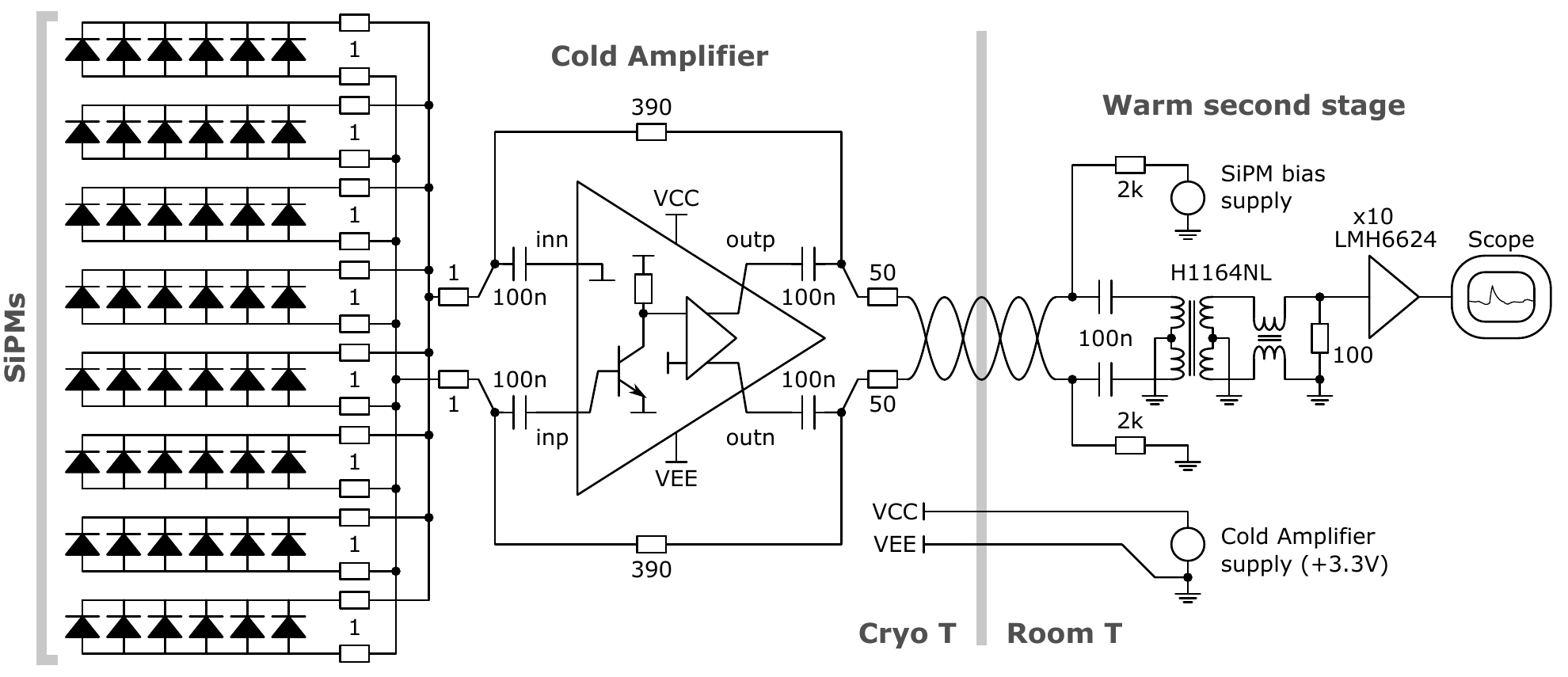}
\caption{\label{fig:GangingSchemeParWarm} Scheme of the readout chain used for the measurements presented in this paper. The detailed schematic of the amplifier is shown in figure \ref{fig:AmpFull}. The DUNE FD1-HD photon detector system will follow the same general scheme, with the warm second stage replaced by a dedicated system.}
\end{figure}}
Figure \ref{fig:GangingSchemeParWarm} shows the scheme of the readout chain for one supercell used for all tests presented in this paper.
The SiPMs are connected in parallel in groups of six. Eight groups of six are then ganged through pairs of \mbox{1 $\Omega$} resistors. All 48 SiPMs are therefore essentially connected in parallel.
The common SiPM anodes and cathodes are AC-coupled to the inputs of the inverting pseudo-differential amplifier, which drives the amplified signals over a 100 $\Omega$ differential pair.
The presence of \mbox{1 $\Omega$} resistors helps to ensure stability of the feedback loop at high frequency.
At the warm side, a H1164NL transformer is used for differential to single-ended conversion, followed by the 100 $\Omega$ termination resistor.
Signals are then further amplified by a LMH6624 operational amplifier with a gain of 10 and acquired by the oscilloscope (Rohde\&Schwarz RTE1054).
Bias for the SiPMs is provided in DC on the same differential pair used in AC for signal readout.

The readout of the DUNE FD1-HD photon detection system will follow the same general scheme, but the warm side will be replaced by a dedicated system.
Power supplies $V_{CC}$ and $V_{EE}$ for the amplifier, as well as the ground reference, will be shared among the four channels in the same module.
The minimum total number of individual wires needed for a photon detection module is then 11 (ground, $V_{CC}$, $V_{EE}$, and four differential pairs), all carried by the same multi-conductor shielded cable.

\section{Amplifier design}
\label{sec:AmpDesign}

\begin{figure}[b]
\centering
\includegraphics[scale=1]{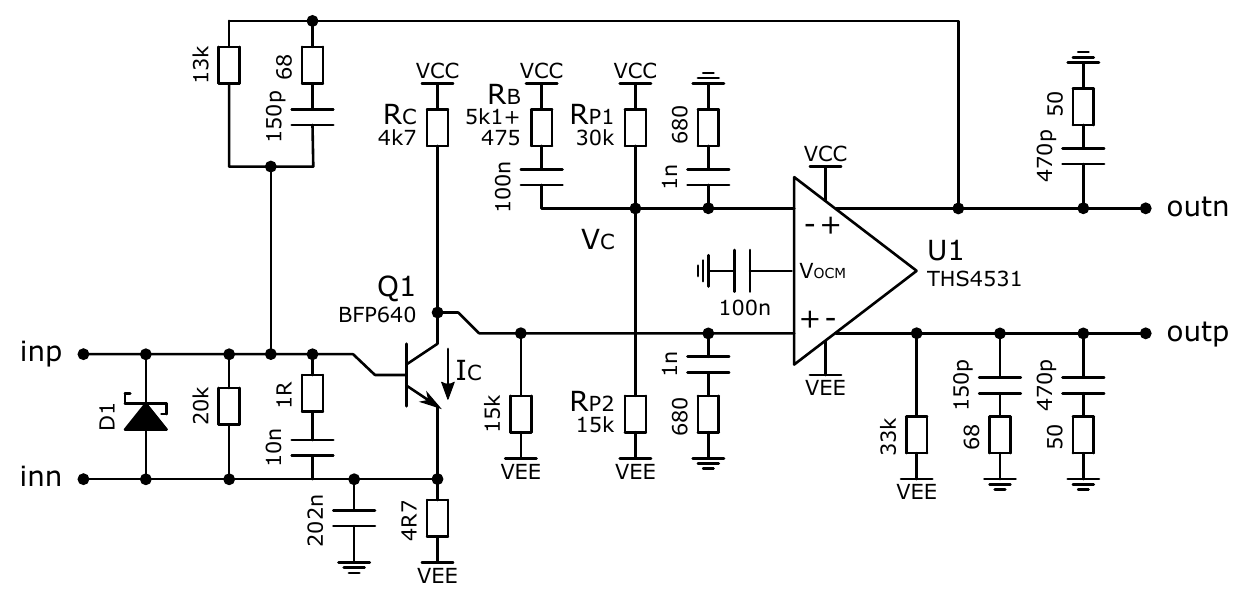}
\caption{\label{fig:AmpFull} Detailed schematic of the cold amplifier (the triangular block of figure \ref{fig:GangingSchemeParWarm}), designed to operate in liquid Argon.}
\end{figure}
Figure \ref{fig:AmpFull} shows the schematic of the cold amplifier.
The design is the evolution of an already published work \cite{ColdAmp1}.
It operates with a single power supply \mbox{$V_{CC}=$ 3.3 V}. The supply current returns on \mbox{$V_{EE}=$ 0 V}. Both supplies are filtered to ground with \mbox{100 nF} caramic capacitors, not shown in the schematic. The ground connection is used as a reference and shielding potential, and does not carry DC currents.
The input device $Q_1$ is a Infineon BFP640ESD silicon-germanium (SiGe) heterojunction bipolar transistor (HBT), whose construction gives low base spreading resistance, hence low voltage noise even at low bias currents, and capability to operate at cryogenic temperature \cite{Cressler}.
A SB01-15C Schottky diode $D_1$ protects its base terminal from accidental reverse biasing or spikes that might propagate through the AC coupling capacitors.
The 10~nF and 1~$\Omega$ at the input, mounted close to $Q_1$, help stability by shorting its base and emitter at high frequency, preventing possible issues due to stray inductances. Their effect at signal frequencies with all SiPMs connected is negligible.
The emitter of $Q_1$ is bypassed to ground with two \mbox{100 nF} and two \mbox{1 nF} capacitors, represented in the schematic as a single \mbox{202 nF} capacitor to ground.
The capacitors are placed as close as possible to the emitter pins, again to minimize possible high frequency instability of $Q_1$ due to parasitic inductances.
The two inputs of the circuit ``inp'' and ``inn'' are connected to the base and emitter of $Q_1$, respectively. The latter is bypassed to ground. While it is still true that the design will amplify the voltage difference between the positive and negative inputs, the input impedances at signal frequencies are clearly different. As such, this has to be considered a pseudo-differential configuration.

The HBT is followed by a second gain stage based on a Texas Instruments THS4531A fully differential operational amplifier $U_1$, based on a BiCMOS technology and found to operate satisfactorily at cryogenic temperature.
It is operated at 60\% of its maximum rating of \mbox{5.5 V}, which greatly reduces possible concerns related with aging due to hot carrier effects \cite{RadekaAndCo}.
All resistors are precision thin metal film with temperature coefficients in the \mbox{$\pm$25 ppm/$^{\circ}$C} range (Vishay CPF, Panasonic ERA, Vishay MMA).
Capacitors are multi-layer ceramics with C0G dielectric (TDK C, Murata GRM).
Those that need to withstand the SiPM bias voltage (up to $\sim$50 V) are rated for \mbox{100 V} DC.

The voltage divider formed by $R_{P1}$ and $R_{P2}$ sets the voltage \mbox{$V_C = V_{CC}/3 =$ 1.1 V}  at the negative input of $U_1$. The feedback loop ensures that this DC voltage is carried over to the other input of $U_1$, where the collector of $Q_1$ is connected.
At cryogenic temperature, the input node ``inp'' sits at \mbox{$\sim$1 V}, the $V_{BE}$ of $Q_1$. A \mbox{20 k$\Omega$} resistor is used to bleed a small current between base and emitter, which develops a voltage across the \mbox{13 k$\Omega$} feedback resistor and brings the output node ``outn'' to \mbox{$\sim$1.65 V}, approximately equal to $V_{CC}/2$.
Since $V_{OCM}$ is biased to $V_{CC}/2$ by \mbox{$\sim$100 k$\Omega$} resistors internal to the THS4531A, the other output ``outp'' is at \mbox{$\sim$1.65 V} as well.
The bias current for $Q_1$ is set by resistor $R_C$ to \mbox{$I_C=$ 0.4 mA}.
The DC current drawn by the THS4531A at cryogenic temperature is about \mbox{0.2 mA}.
Approximately \mbox{0.1 mA} are used by the voltage dividers in the circuit.
The total DC current drawn by the amplifier is about \mbox{0.7 mA}, for a total power consumption of \mbox{2.4 mW} per channel.

At \mbox{$I_C=$ 0.4 mA}, the HBT exhibits a voltage white noise of \mbox{0.37 nV/$\surd$Hz}, and 1/f noise contributions can be neglected above a few kHz \cite{ColdAmp1}.
The base current is about 1 $\mu$A.
Due to the low input impedance presented by the group of ganged SiPMs, the weight of the current noise is negligible compared to the voltage noise.
The HBT at LAr temperature (\mbox{87 K}) has a transconductance of \mbox{50 mA/V} and operates at an AC gain of $\sim$30, enough to make the noise of $U_1$ negligible.
The purpose of the articulate voltage divider between $Q_1$ and $U_1$ (including the precise value of $R_B$, chosen so that $R_B || R_{P1} = R_C$) is to keep a balanced impedance at the inputs of $U_1$, which improves the rejection of noise and interference from the power supplies.
For the same reason the output common mode voltage input $V_{OCM}$ of $U_1$ is filtered to ground with \mbox{100 nF}.
Care was taken to make sure that the loads at the two outputs of the THS4531A were balanced over the entire frequency range.
Additional 50 $\Omega$ resistors were used to load the output stage of the THS4531A at high frequency, lowering its open-loop gain and increasing the phase margin to avoid low amplitude oscillations that could occur at cryogenic temperature.
With components as shown, the useful linear differential swing at the output of the amplifier is about \mbox{1.5 - 1.6 V}.
Beyond this range, soft saturation occurs and distorts the signal shape.

The main characteristics of the amplifier can be compared to those of a similar system based on the LMH6629 opamp \cite{DarkSideAmp1}.
The input-referred series noise of the two designs is comparable: the LMH6629 has the noise of a 20 $\Omega$ resistor \cite{DarkSideAmp1}, which at cryogenic temperature amounts to \mbox{0.3 nV/$\surd$Hz}, compared to the \mbox{0.37 nV/$\surd$Hz} of the design presented here. At the same time, the power consumption of the design described here, \mbox{2.4 mW}, is more than an order of magnitude smaller than the \mbox{50 mW} per amplifier of the LMH6629 \cite{DarkSideAmp1}.
Comparing bandwidth, the gain-bandwidth product of the LMH6629 is about 10 GHz at cryogenic temperature \cite{DarkSideAmp2}. The design presented here has a gain-bandwidth product of around 1.5 GHz, given by the gain-bandwidth product of the THS4531 (about 50 MHz at cryogenic temperature, roughly twice than at room temperature \cite{ColdAmp1}), multiplied by the gain of the first stage ($\sim$30).
This design offers a higher flexibility compared to any solution based on a single opamp, since the gain-bandwidth product can be tuned by changing the gain of the first stage. In any case, the closed-loop bandwidth cannot be larger than approximately 50 MHz, the frequency of the second pole of the THS4531 at cold, which corresponds to a minimum rise time of $\sim$10 ns.

\FloatBarrier
\section{Signal shape and gain}
\label{sec:Signals}

\begin{figure}[b]
\centering
\includegraphics[scale=0.75]{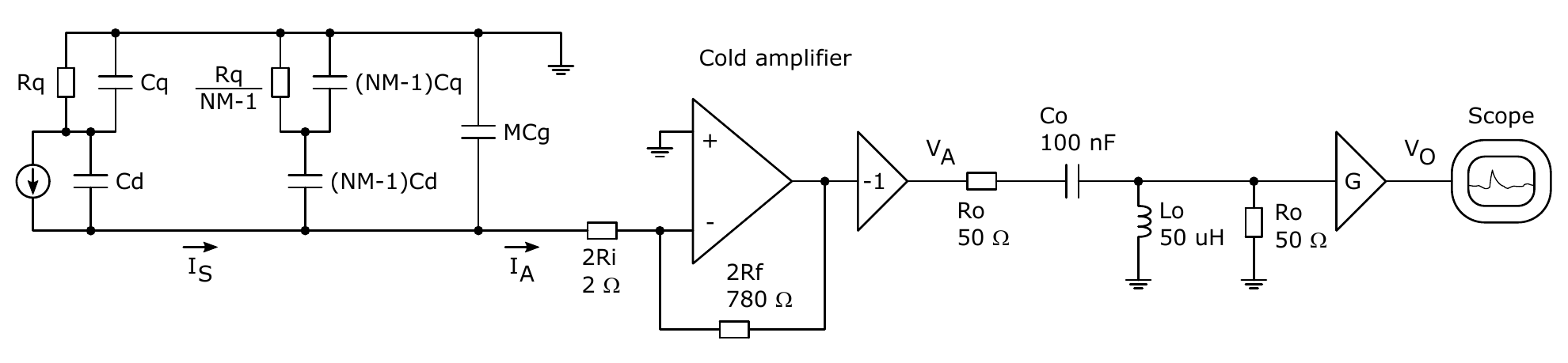}
\caption{\label{fig:EquivalentSignal} Equivalent simplified schematic of the readout chain, for the purpose of calculating the expected signal shape.}
\end{figure}

Figure \ref{fig:EquivalentSignal} shows the single-ended equivalent circuit that can be used to estimate the expected signal shape in response to the current signal from a single SiPM cell (a single photoelectron signal).
Here $N$ is the number of cells in a SiPM, while $M=48$ is the number of SiPMs ganged in parallel.
The 1 $\Omega$ resistors in series with each group of 6 SiPM were neglected, while the two 1 $\Omega$ resistors at the input of the amplifier are lumped in the \mbox{$2 R_i=$ 2 $\Omega$} input resistance.
Assuming negligible $C_q$ and $C_g$, and with values from table \ref{tab:SiPMs}, the low frequency impedance of 48 SiPMs in parallel is about \mbox{70 nF} for SiPM model 1 and \mbox{100 nF} for SiPM model 2.
At signal frequencies, above $s \sim 1/\tau_S$, the equivalent resistance of 48 SiPMs is about 7 $\Omega$ for both SiPM models, which is not negligible compared with the 2 $\Omega$ input resistance of the amplifier.
While the signal current from one cell can still be modeled with equation \ref{eq:SiPMcurrent1} or \ref{eq:SiPMcurrent2}, it will be divided between $2 R_i$ and the impedance of $M$ SiPMs in parallel, that is $Z_S/M$, with $Z_S$ given by equation \ref{eq:SiPMimpedance1} or \ref{eq:SiPMimpedance2}.

The current that passes through $2 R_i$ is amplified by the transimpedance gain of the amplifier, \mbox{$2 R_f=$ 780 $\Omega$}, where the factor $2$ comes from the differential configuration.
The bandwidth limit of the output signals can be represented as a pole with time constant \mbox{$\tau_{B} \simeq$ 40 ns}, corresponding to a closed-loop bandwidth of approximately \mbox{4 MHz}.
An ideal inverting amplifier with gain $-1$ was added to the equivalent schematic to reflect the correct polarity received by the oscilloscope.
In the domain of the complex frequency $s$, the signal at the node $V_A$, after the amplifier and before the AC-coupling, can be written as
\begin{equation}
    \tilde{V}_A(s) =  \tilde{I}_A(s) \left( \frac{2 R_f}{1+s\tau_{B}} \right) =
    \tilde{I}_S(s) \left(\frac{Z_{S}(s)/M}{Z_{S}(s)/M+2 R_i}\right) \left( \frac{2 R_f}{1+s\tau_{B}} \right),
     \label{eq:AmpVoltage}
\end{equation}
where $\tilde{I}_S(s)$ is the Laplace transform of equation \ref{eq:SiPMcurrent1} or \ref{eq:SiPMcurrent2}.

At the output of the amplifier, the signal is AC coupled to the transformer, whose load is the line termination resistor, here modeled as 50 $\Omega$ single-ended.
As shown in figure \ref{fig:GangingSchemeParWarm}, the H1164NL is actually composed of a transformer followed by a common mode choke. The values of inductance are not clearly specified.
While an ideal transformer has infinite (hence negligible) self-inductance, a real transformer has a finite self-inductance, often called magnetization inductance, which can be modeled in parallel to the primary winding of an ideal transformer.
With a reasonable degree of approximation, the H1164NL can be modeled as a single inductor of inductance $L_o$, where the value \mbox{$L_o=$ 50 $\mu$H} was extracted as a free parameter to match the shape of the measured waveforms presented later.
Of the two sets of \mbox{100 nF} AC-coupling capacitors, only the warm side was considered, which became just one capacitor in the equivalent single-ended representation.
The AC-coupling capacitors at the cold side can be neglected, since they are inside the feedback loop of the amplifier, hence their effective value is made larger by a factor equal to the loop gain.
Adding also a gain $G=10$ from the second stage, the signal that feeds the oscilloscope is given by
       \begin{equation}
       \begin{split}
      \tilde{V}_O(s) &=
      G \tilde{V}_A(s)
      \left(\frac{s^2 L_o C_o}{1+s (L_o/R_o+R_o C_o) + 2 s^2 L_o C_o}\right)\\
      &=    G
      \tilde{I}_S(s) \left(\frac{Z_{S}(s)/M}{Z_{S}(s)/M+2 R_i}\right) \left( \frac{2 R_f}{1+s\tau_{B}} \right)
      \left(\frac{s^2 L_o C_o}{1+s (L_o/R_o+R_o C_o) + 2 s^2 L_o C_o}\right).
      \end{split}
         \label{eq:ScopeVoltage}
    \end{equation}
The inverse Laplace transform of equation \ref{eq:ScopeVoltage}, whose analytical expression is not particularly enlightening, gives the expected response in time domain.

\begin{figure}[t]
\centering
\includegraphics[scale=0.5]{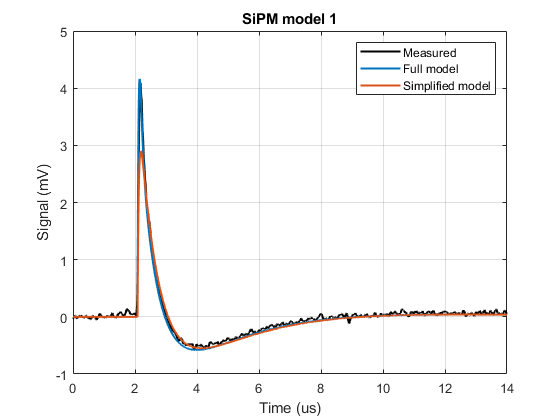}
\includegraphics[scale=0.5]{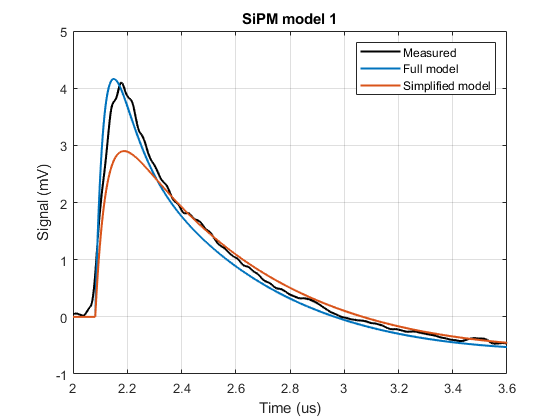}
\includegraphics[scale=0.5]{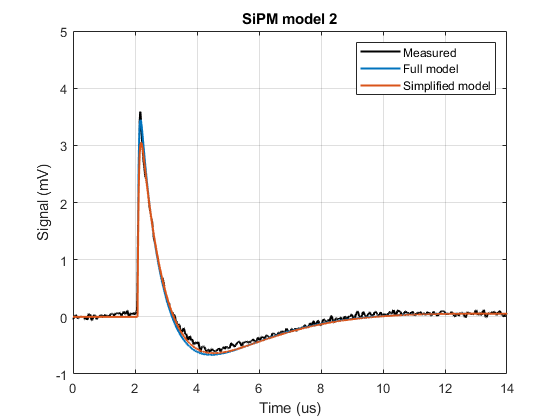}
\includegraphics[scale=0.5]{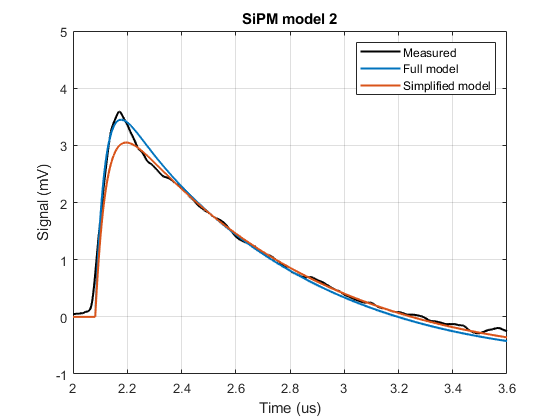}
\caption{\label{fig:Signals1pe} Average single photon signals in black, acquired by the oscilloscope with the setup of figure \ref{fig:GangingSchemeParWarm}. SiPM model 1 above, SiPM model 2 below, on different time scales. Both SiPMs are operated at the overvoltage corresponding to 45\% PDE. The calculated waveforms according to equation \ref{eq:ScopeVoltage} are shown in blue (``full model'', with estimated values for $C_q$ and $C_g$), and orange (``simplified model'', neglecting $C_q$ and $C_g$).}
\end{figure}

Figure \ref{fig:Signals1pe} shows, for both SiPM models, the measured signals as seen at the oscilloscope in response to single photons from a pulsed LED.
In both cases, the signals were acquired at the overvoltage that corresponds to 45\% PDE.
The signals were averaged over $\sim 1000$ acquired waveforms, which were indentified as single photoelectron signals according to the distributions that will be shown in section \ref{sec:Noise}.
Along with the measured signals, figure \ref{fig:Signals1pe} shows the waveforms calculated from the inverse Laplace transform of equation \ref{eq:ScopeVoltage}, with SiPM parameters taken from table \ref{tab:SiPMs}.
The blue curves in figure \ref{fig:Signals1pe} are calculated using expressions \ref{eq:SiPMcurrent1} and \ref{eq:SiPMimpedance1} for $I_S$ and $Z_S$, which do not neglect $C_q$ and $C_g$.
To obtain a good match, $C_q$ was set to \mbox{50 fF} for SiPM model 1 (about 0.25 $C_d$), and \mbox{20 fF} for SiPM model 2 (about 0.15 $C_d$).
For both SiPM models $C_g$ was set to \mbox{10 pF}.
The blue waveforms successfully reproduce all the features of the measured signal.
The orange curves in figure \ref{fig:Signals1pe} are calculated according to the simplified expressions \ref{eq:SiPMcurrent2} and \ref{eq:SiPMimpedance2} for $I_S$ and $Z_S$, which neglect $C_q$ and $C_g$.
The simplified model matches reasonably well the waveform of SiPM model 2, less so for the first $\sim$200 ns of the signal of SiPM model 1 (between 2.1 $\mu$s and 2.3 $\mu$s of the time axis of figure \ref{fig:Signals1pe}), which has a larger $C_q$.
The orange and blue curves are essentially equivalent on time scales of \mbox{200 ns} or more, from 2.3 $\mu$s onwards.

For both SiPM models, the 10\% to 90\% rise time of the signals is \mbox{$\sim$70 ns}.
The 90\% to 10\% fall time is \mbox{580 ns} for SiPM model 1 and \mbox{780 ns} for SiPM model 2, which is about half than what would be expected from SiPM signals alone, given their time constant $\tau_S$.
This is of course due to the AC-coupling at the output of the amplifier, which makes the signals bipolar, with an undershoot following each signal, with a time constant of a few $\mu$s.
The shortening of the tail implies that a fraction of the total charge carried by each signal cannot be recovered by integrating the signals in time, and needs to be accounted for when measuring the SiPM gain.
The inconvenience of having a bipolar signal shape is outweighted by the practical advantage of using the same wires for SiPM biasing (DC) and signal readout (AC).
The measured peak amplitude at the oscilloscope is \mbox{3.5-4 mV} for both SiPM models at this overvoltage. This corresponds to a \mbox{0.7-0.8 mV} differential signal at the output of the amplifier (node $V_A$), which gives a useful dynamic range of about 2000 photoelectrons before saturation occurs.

\section{Noise and photon counting resolution}
\label{sec:Noise}

\begin{figure}[b]
\centering
\includegraphics[scale=0.75]{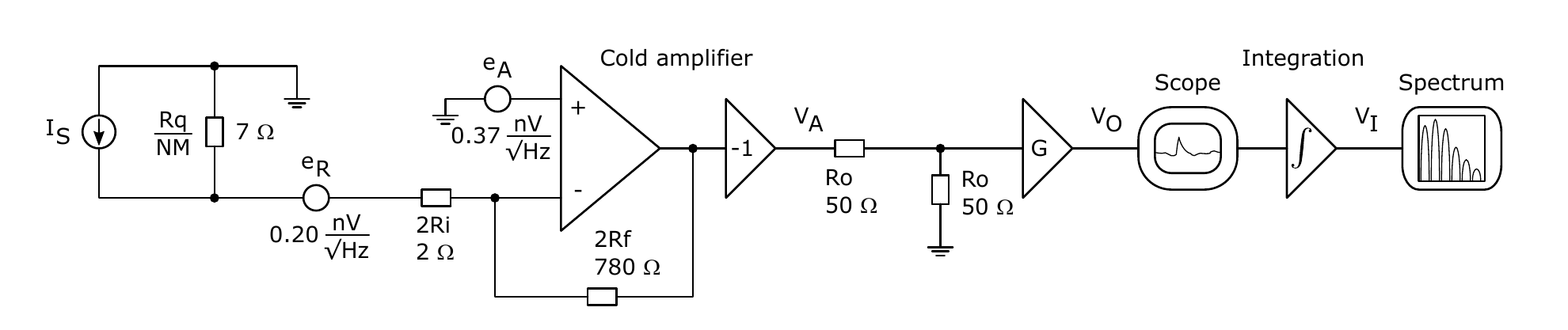}
\caption{\label{fig:EquivalentNoise} Equivalent simplified schematic of the readout chain, for the purpose of calculating the expected signal to noise ratio of the integrated spectra.}
\end{figure}

To measure the charge carried by each signal, signals are integrated over a finite time window, until approximately the first zero crossing due to the AC-coupling, which occurs after about 1 $\mu$s.
Figure \ref{fig:EquivalentNoise} shows the equivalent circuit used to predict the effect of the fundamental noise sources on the resolution of the photoelectron spectra.
The source impedance of the $M=48$ SiPMs in parallel was simplified to its resistance at signal frequencies $R_q/NM$.
The AC-coupling at the output of the amplifier was neglected (infinite $L_o$ and $C_o$).

The dominant contributor to noise is the voltage white noise of the amplifier \mbox{$e_A=$ 0.37 nV/$\surd$Hz}, as was already stated in section \ref{sec:AmpDesign}.
Since $e_A$ is low, the thermal noise of the total SiPM quenching resistance \mbox{$R_q/NM\simeq$ 7 $\Omega$} and of the amplifier input resistance \mbox{$2 R_i=$ 2 $\Omega$} are not negligible even at cryogenic temperature, where they contribute with \mbox{$\sqrt{4kTR_q/NM}=$ 0.18 nV/$\surd$Hz} and \mbox{$\sqrt{4kT2R_i}=$ 0.10 nV/$\surd$Hz} respectively. These two contributions can be summed in quadrature and represented as a single noise source \mbox{$e_R=$ 0.20 nV/$\surd$Hz}.
The two noise generators $e_A$ and $e_R$ have slightly different gains, but the difference can be neglected, since the closed loop gain of the cold amplifier is much larger than 1.

Aside for a normalization factor equal to $T_I$, a moving integration over a time window $T_I$ is equivalent to a moving average filter with the same window.
Such filter has a -3dB cutoff frequency $f_{-3dB}=0.443/T_I$.
Although the behaviour in the stop band is somewhat different, it can be well approximated by a low pass filter with transfer function $\tau_I/(1+s\tau_I)$, where $\tau_I=1/(2 \pi f_{-3dB}) = T_I/2.78$, which is better suited for analysis in the complex frequency domain.
A 1 $\mu$s integration window then corresponds to a low pass filter with \mbox{$\tau_I\simeq$ 360 ns}.
With this integration range the bandwidth limit previously modeled with $\tau_B$ can be neglected.
Since it represents an integration in time, the transfer function of the filter has dimensions of time, and the quantities expressed below (equations \ref{eq:IntNoise} to \ref{eq:IntVoltageMax}) are expressed in V$\cdot$s.
The noise at the node $V_I$ is given by
\begin{equation}
   v_n(s) = \sqrt{e_A^2+e_R^2} \left(\frac{2 R_f}{2 R_i+R_q/NM}\right) \left(\frac{G}{2}\right)\left(\frac{\tau_I}{1+s\tau_I}\right).
\label{eq:IntNoise}
\end{equation}
The RMS value can be calculated by integrating its squared amplitude over all frequencies and then taking the square root:
\begin{equation}
\begin{split}
   v_{RMS} =
   \sqrt{\int_0^{\infty} \left|v_n(s) \right|^2 \frac{d \omega}{2 \pi}}
   & \simeq \sqrt{e_A^2+e_R^2} \left(\frac{G R_f}{2 R_i+R_q/NM}\right)
   \sqrt{\int_0^{\infty} \frac{\tau_I^2}{1+\omega^2\tau_I^2} \frac{d \omega}{2 \pi}}\\
   & =\sqrt{e_A^2+e_R^2} \left( \frac{G R_f}{2 R_i+R_q/NM} \right) \frac{\sqrt{\tau_I}}{2}.
   \end{split}
   \label{eq:IntNoiseRMS}
\end{equation}
Since the only SiPM parameter that enters equation \ref{eq:IntNoiseRMS} is the total quenching resistance per SiPM, $R_q/N$, which is about the same value for both SiPM models, the predicted integrated RMS noise is the same.

The integrated noise should be compared with the integral of the amplified single photon signal.
Following the same equivalent circuit of figure \ref{fig:EquivalentNoise} the integrated waveform at node $V_I$ is given by
\begin{equation}
\begin{split}
    \tilde{V}_I(s) &= 
      \tilde{I}_S(s) \left(\frac{R_q/NM}{2 R_i+R_q/NM}\right) \left(2 R_f\frac{G}{2}\right)
       \left(\frac{\tau_I}{1+s\tau_I}\right)\\
       & =     G R_f
     \frac{V_{OV}}{R_q}
       \left(\frac{R_q/NM}{2 R_i+R_q/NM}\right)
        \left(\frac{\tau_S}{1+s\tau_S}\right)
       \left(\frac{\tau_I}{1+s\tau_I}\right),
       \end{split}
       \label{eq:IntVoltage}
\end{equation}
where we used the Laplace transform of equation \ref{eq:SiPMcurrent2} for $\tilde{I}_S$.
Since we are modeling the integration with a moving low pass filter, the integral value should coincide with the maximum of the inverse Laplace transform of equation \ref{eq:IntVoltage}, which corresponds to the case where the integration window is optimally centered on the waveform.
The maximum can be expressed by
\begin{equation}
   V_{MAX} = 
       G R_f
     \frac{V_{OV}}{R_q}
       \left(\frac{R_q/NM}{2 R_i+R_q/NM}\right) \alpha\left(\tau_S, \tau_I \right),
       \label{eq:IntVoltageMax}
\end{equation}
where $\alpha \left(\tau_S,\tau_I\right)$ is a factor with dimensions of time resulting from the peak amplitude of a double exponential:
\begin{equation}
     \alpha \left(\tau_S,\tau_I\right) = \frac{\tau_S \tau_I}{\tau_S-\tau_I}
     \left[ \left( \frac{\tau_I}{\tau_S} \right)^{\frac{\tau_I}{\tau_S-\tau_I}}-
     \left( \frac{\tau_I}{\tau_S} \right)^{\frac{\tau_S}{\tau_S-\tau_I}}\right].
     \label{eq:alpha}
\end{equation}

The signal to noise ratio is defined as the single photoelectron gain divided by the root-mean-square noise of the integrated baseline, and can now be estimated by dividing equation \ref{eq:IntVoltageMax} by \ref{eq:IntNoiseRMS}:
\begin{equation}
    \frac{V_{MAX}}{v_{RMS}} = \frac{2 \alpha (\tau_S, \tau_I) V_{OV}}{N M \sqrt{\tau_I} \sqrt{e_A^2+e_R^2} }.
    \label{eq:SignalToNoise}
\end{equation}
The signal to noise estimate expressed by equation \ref{eq:SignalToNoise} is directly proportional to the overvoltage $V_{OV}$.
Operating at high overvoltage increases the resolution of the photoelectron spectra, but increases the crosstalk and afterpulse probability, which is undesirable.
The total voltage noise $\sqrt{e_A^2+e_R^2}$ appears at the denominator, which clearly emphasizes the advantage of minimizing this term.
The signal to noise is also inversely proportional to the total number of SiPMs $M$, although this is true only as long as $e_R$ is smaller than $e_A$, since $e_R \sim \sqrt{R_q/NM}$. If $e_A$ was negligible compared with $e_R$ (which is not the case here), then the signal to noise would scale with the number is SiPMs as $1/\sqrt{M}$.
Note that equation \ref{eq:SignalToNoise} is valid as long as the closed loop gain of the amplifier is much larger than 1, i.e. $R_f \gg 2 R_i+R_q/NM$, which led us to treat $e_A$ and $e_R$ on equal grounds in equation \ref{eq:IntNoise}.

\begin{figure}[t]
\centering
\includegraphics[scale=0.38]{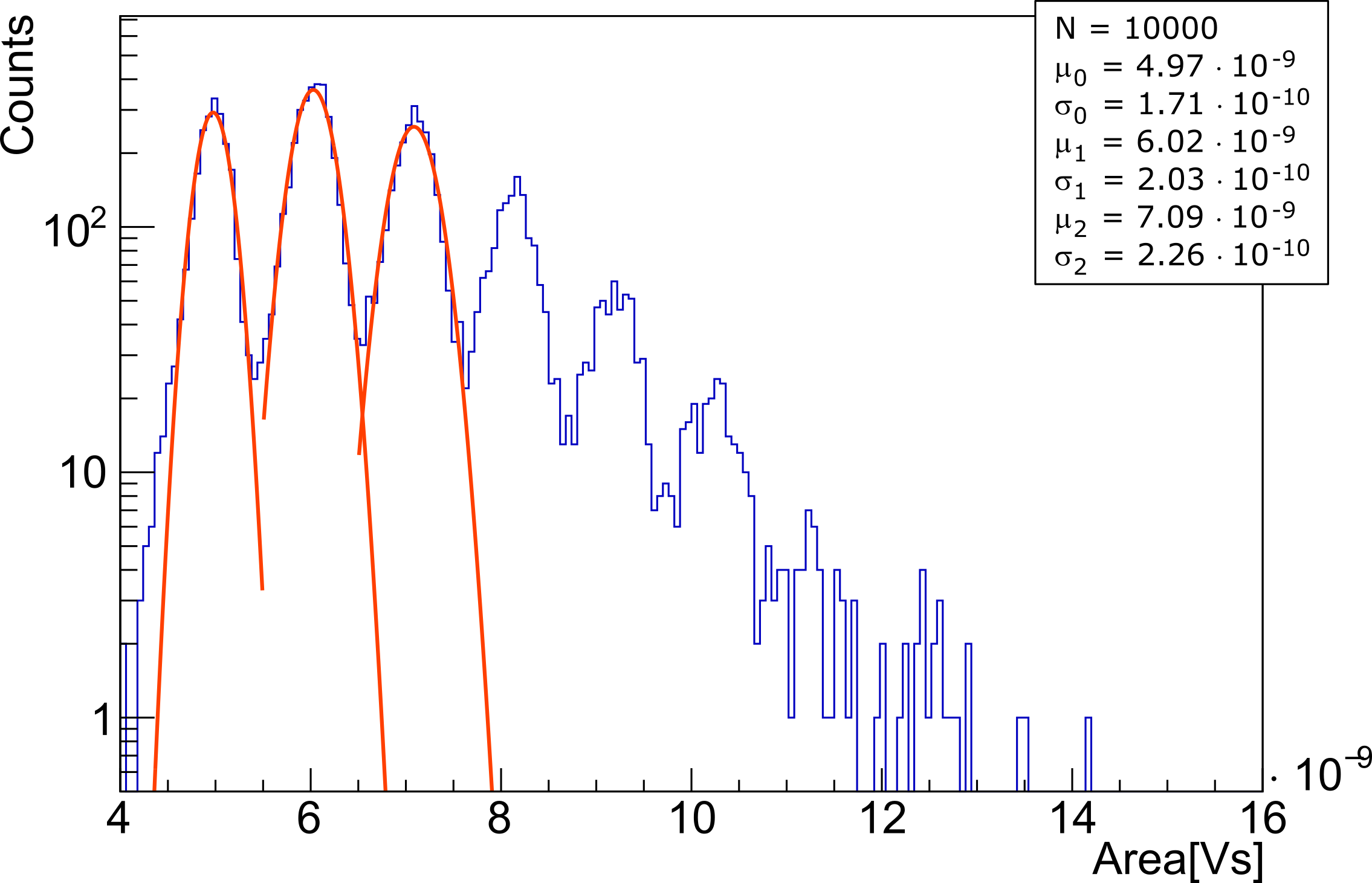}
\includegraphics[scale=0.38]{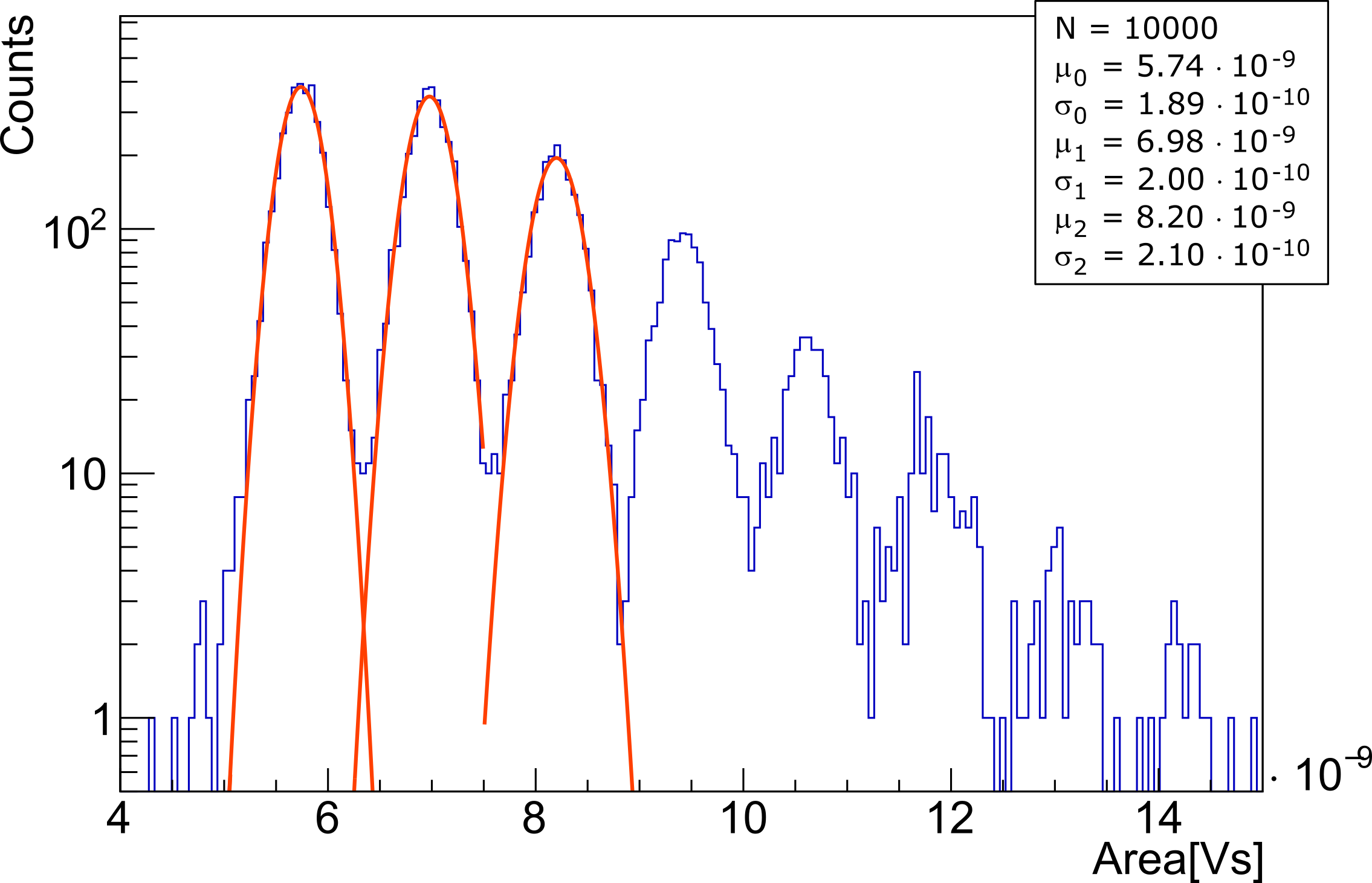}

\vspace{0.2cm}

\includegraphics[scale=0.38]{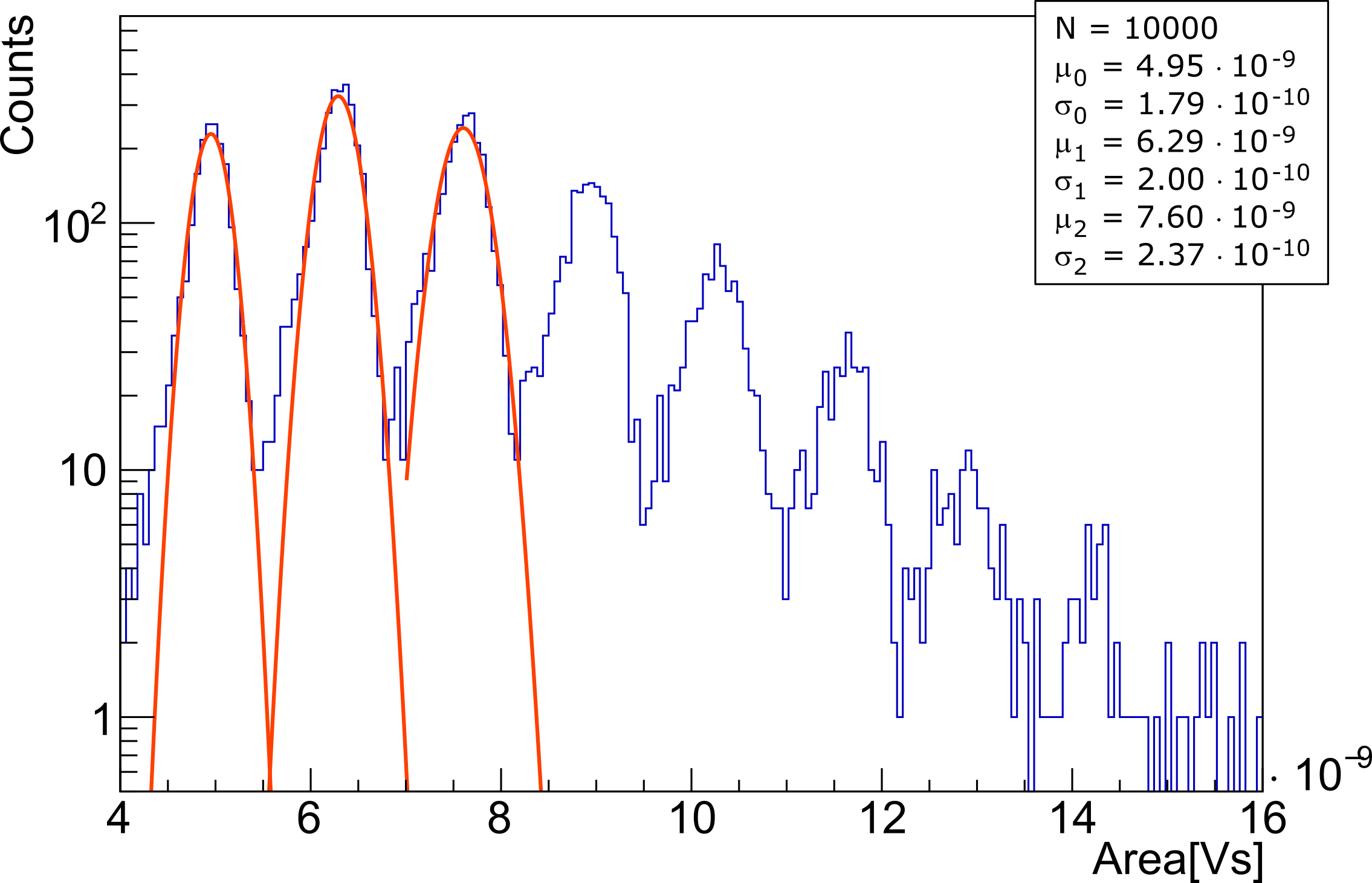}
\includegraphics[scale=0.38]{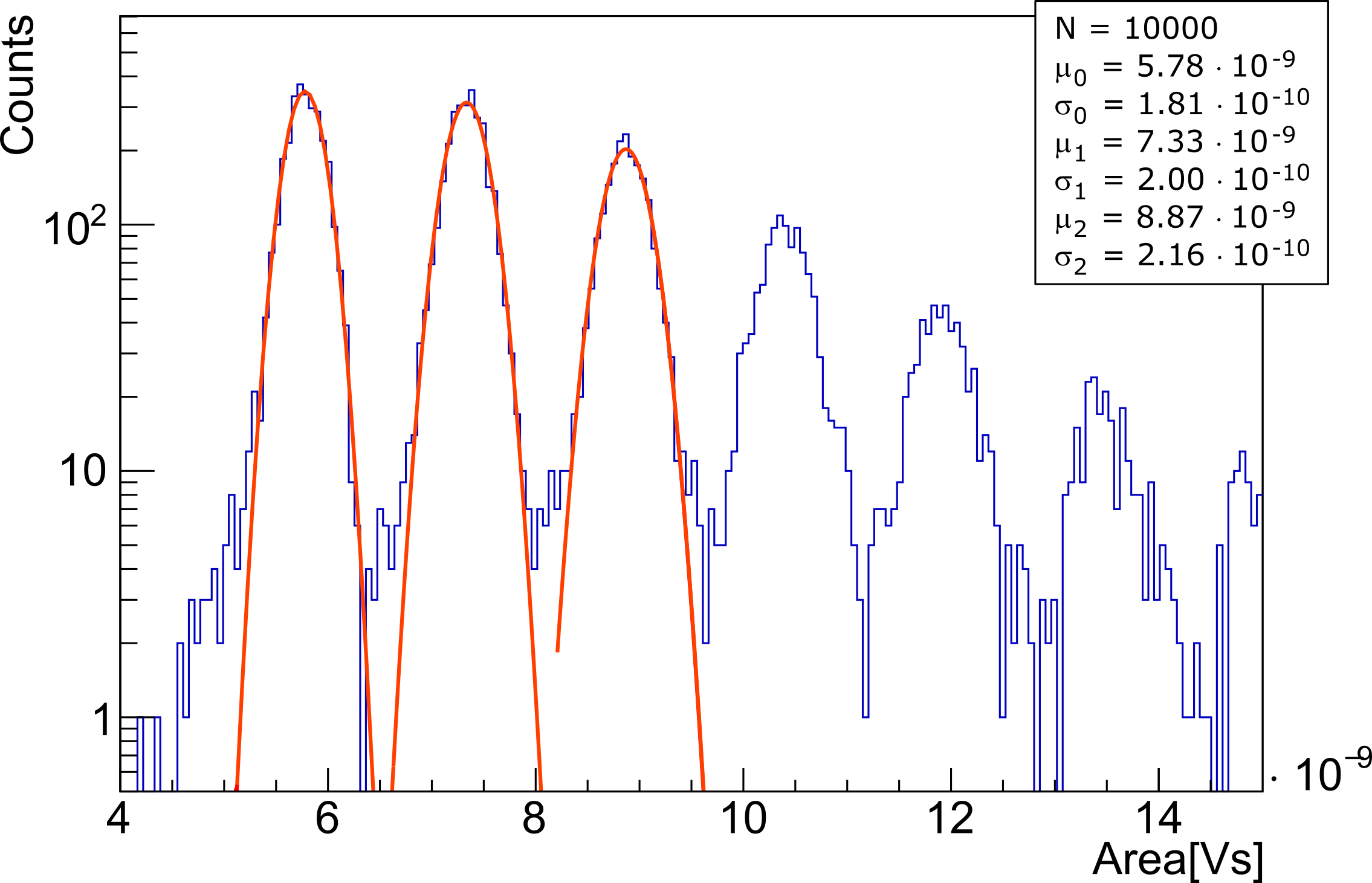}

\vspace{0.2cm}

\includegraphics[scale=0.38]{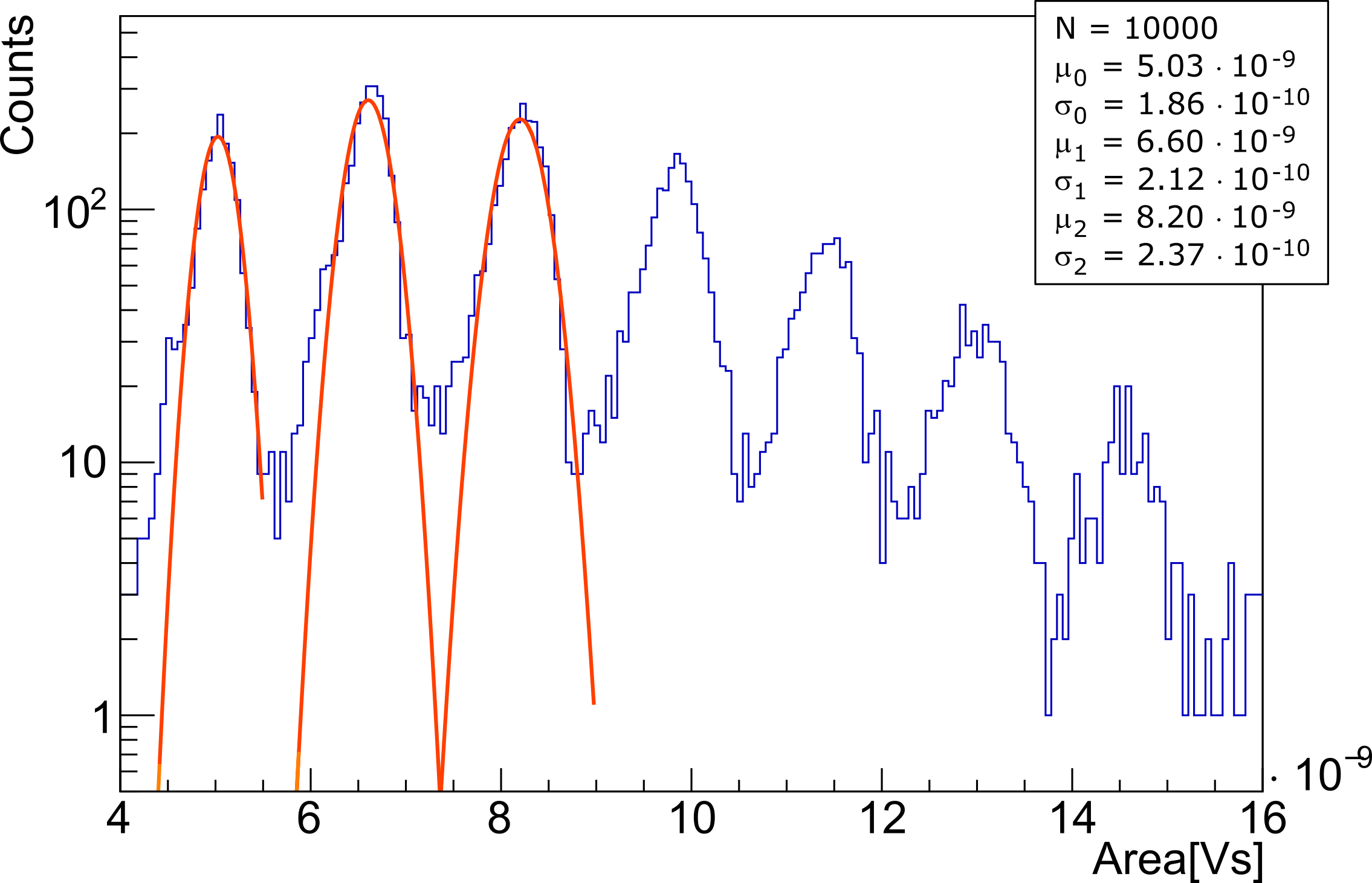}
\includegraphics[scale=0.38]{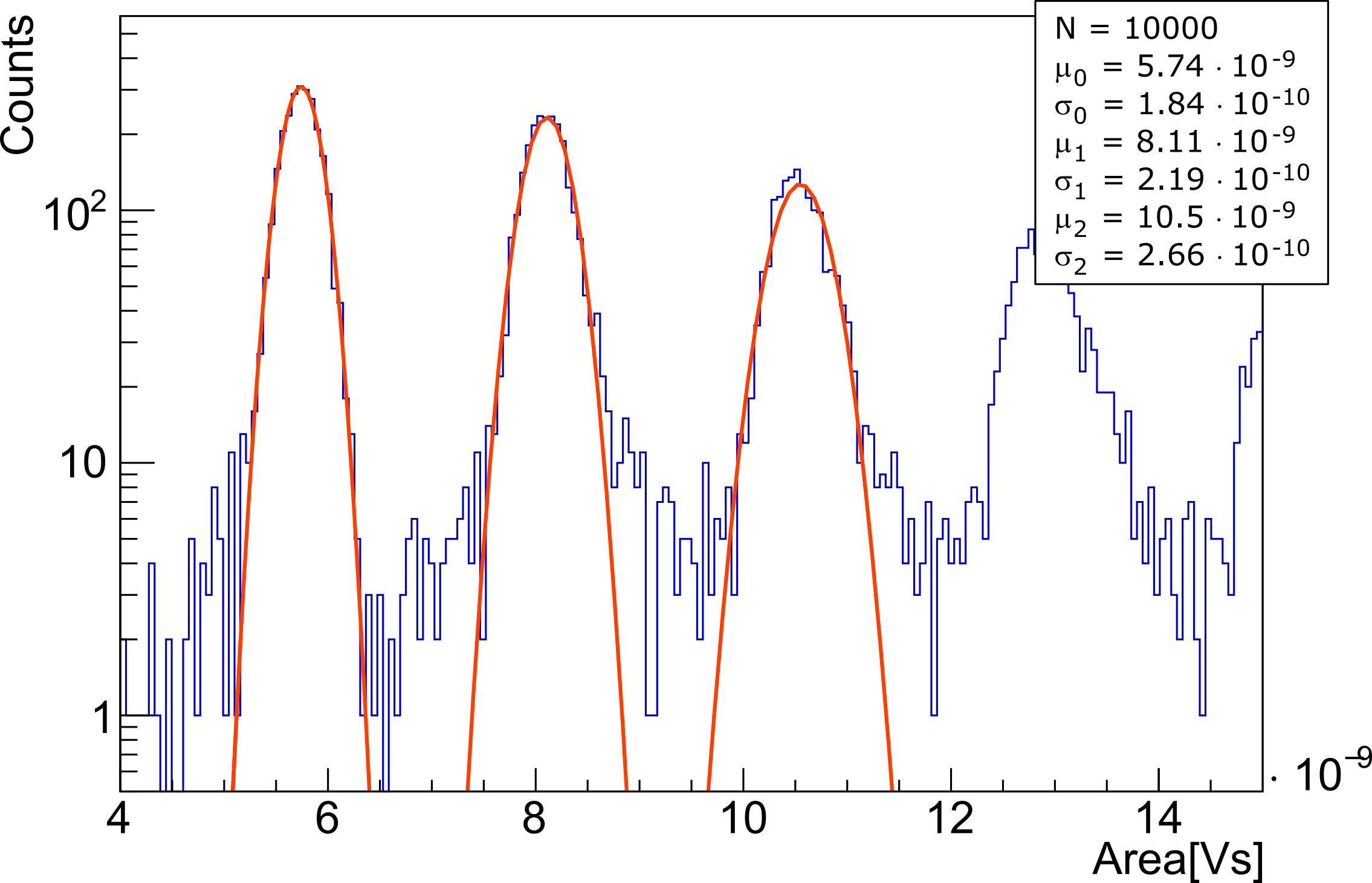}
\caption{\label{fig:Spectra} Single photoelectron spectra obtained reading out 48 \mbox{$6\times6$ mm$^2$} SiPMs with the cold amplifier described in this paper, as in the scheme of figure \ref{fig:GangingSchemeParWarm}. SiPM model 1 is shown on the left, SiPM model 2 on the right. The three rows are measured at different overvoltages, corresponding to PDE 40\% (top), 45\% (middle), 50\% (bottom).
For each spectrum, the peaks corresponding to 0, 1, 2 photoelectrons were fitted with Gaussian functions. The fit parameters are shown in the plot inlays.}
\end{figure}

The dependency of equation \ref{eq:SignalToNoise} on the integration time $\tau_{I}$ depends on the factor $\alpha (\tau_S, \tau_I)/\sqrt{\tau_I}$, which predicts a better signal to noise ratio for larger values of $\tau_I$.
In practice, the width of the integration window can be tuned in the \mbox{600 ns - 1 $\mu$s} range while looking for the best signal to noise ratio.
Larger values, up to the zero crossing seen in figure \ref{fig:Signals1pe}, integrate more signal, but might show higher sensitivity to low frequency noise or environmental interference.
Even without the AC-coupling at the receiving end of the amplified signals, diminishing returns would come from integrating over longer times, as SiPM afterpulses tend to appear.
An integration window of 1 $\mu$s (\mbox{$\tau_I =$ 360 ns}) and a fall time of SiPM signals of \mbox{$\tau \simeq$ 600 ns} give \mbox{$\alpha \simeq$ 167 ns}.
Equation \ref{eq:SignalToNoise} then predicts a signal to noise ratio of $\sim 11$ for both SiPM models operated at the overvoltage that corresponds to 45\% PDE.
But actually, this has been obtained neglecting the AC-coupling at the output of the amplifier, since $C_o$ and $L_o$ were omitted from the equivalent schematic of figure \ref{fig:EquivalentNoise}.
We know from section \ref{sec:Signals} that the AC-coupling shortens the signals, resulting in an effective decay time constant of $\sim \tau_S/2$.
But then $\tau_S/2 \simeq \tau_I$. Replacing $\tau_S$ with $\tau_I$ in equation \ref{eq:IntVoltage} gives \mbox{$\alpha = \tau_I e^{-1} =$ 132 ns}.
With this value, equation \ref{eq:SignalToNoise} predicts a signal to noise ratio of $\sim 9$ for both SiPM models operated at the overvoltage corresponding to 45\% PDE.

Figure \ref{fig:Spectra} shows the histograms of the values of waveform integrals, obtained from sets of 10000 waveforms from a pulsed LED set to emit a few photons in average for each pulse.
Integration times were \mbox{800 ns} for SiPM model 1 and \mbox{900 ns} for SiPM model 2.
The peaks corresponding to 0, 1 and 2 photoelectrons were fitted with Gaussian functions, whose parameters are shown in the inset of each plot.
The measured signal to noise is obtained as the difference between the mean of the 0 and 1 photoelectron peaks, divided by the sigma of the 0 photoelectron peak (integrated baseline noise).
At the three overvoltages that correspond to 40\%, 45\% and 50\% PDE, the signal to noise ratios are 6.1, 7.5 and 8.5 for SiPM model 1 and 6.6, 8.6 and 12.9 for SiPM model 2.
This is more than adequate to clearly discern the single photoelectron peaks even at low overvoltage values.
The results are in good agreement with what was calculated from equation \ref{eq:SignalToNoise}, which confirms that the dominant noise sources have been correctly identified, and other contributions are negligible.

Equation \ref{eq:SignalToNoise} can also be used to predict the signal to noise of signals without integration, by setting \mbox{$\tau_I = \tau_B =$ 40 ns}, the bandwidth limit of the amplifier, and \mbox{$\tau_S \simeq$ 360 ns}, to account for the fall time of SiPM signals shortened by the AC coupling. In this case equation \ref{eq:alpha} gives \mbox{$\alpha (\tau_S, \tau_I) \simeq$ 30 ns}, and equation \ref{eq:SignalToNoise} predicts a signal to noise ratio of $\sim 6$ for both SiPM models at 45\% PDE.
The actual signal to noise that was obtained in this case with real data, by building the histogram of signal maxima, was close to 4. This is again in reasonable agreement with what could be estimated with the simplified model.

\section{Status and outlook}

The paper described the cryogenic amplifier designed for the photon detection system of the first far detector (FD1-HD) of the DUNE experiment.
It reads out arrays of 48 \mbox{$6\times6$ mm$^2$} SiPMs connected in parallel, with a rise time below 100 ns, a linear dynamic range up to 2000 photons, and a power consumption of 2.4~mW per channel.
The measured signal to noise ratio after integration is approximately 8 for both SiPM models considered for DUNE FD1-HD, at the overvoltage that corresponds to 45\% photon detection efficiency.
This result is partly affected by the AC coupling at the output, which responds to the practical necessity of sharing the same wires for SiPM bias and signal readout.
Still, the resolution is clearly more than adequate to separate photoelectron peaks and enable photon counting.
The design satisfies the requirements of the DUNE experiment with ample margin, and may be employed in any other experiment needing to read out large SiPM arrays in cryogenic environments.

\FloatBarrier


\begin{thebibliography}{99}

\bibitem{DUNE}
B. Abi, et al.,
\emph{Deep Underground Neutrino Experiment (DUNE), Far Detector Technical Design Report, Volume I: Introduction to DUNE},
arXiv:2002.02967.

\bibitem{FD1}
B. Abi, et al.,
\emph{Deep Underground Neutrino Experiment (DUNE), Far Detector Technical Design Report, Volume IV: Far Detector Single-phase Technology},
arXiv:2002.03010.

\bibitem{Arapuca2}
A.A. Machado, et al.,
\emph{The X-ARAPUCA: an improvement of the ARAPUCA device},
2018 JINST 13 C04026, doi: 10.1088/1748-0221/13/04/C04026.

\bibitem{Arapuca1}
A.A. Machado and E. Segreto,
\emph{ARAPUCA a new device for liquid argon scintillation light detection},
2016 JINST 11 C02004, doi: 10.1088/1748-0221/11/02/C02004.

\bibitem{SiPMmodel1}
S. Cova, et al.,
\emph{Avalanche photodiodes and quenching circuits for single-photon detection},
Applied Optics 35 12 1996 pp.1956-1976, doi: 10.1364/AO.35.001956.

\bibitem{SiPMmodel2}
F. Corsi, et al.,
\emph{Modelling a silicon photomultiplier (SiPM) as a signal source for optimum front-end design},
NIMA 572, 1, 2007, pp.416-418, doi: 10.1016/j.nima.2006.10.219.

\bibitem{SiPMmodel3}
S. Seifert, et al.,
\emph{Simulation of Silicon Photomultiplier Signals},
IEEE TNS, 56 6 2009, pp.3726-3733, doi: 10.1109/TNS.2009.2030728.

\bibitem{SiPMmodel4}
D. Marano, et al.,
\emph{Silicon Photomultipliers Electrical Model Extensive Analytical Analysis},
IEEE TNS, 61 1 2014, pp.23-34, doi: 10.1109/TNS.2013.2283231.

\bibitem{ColdAmp1}
P. Carniti, et al.,
\emph{A low noise and low power cryogenic amplifier for single photoelectron sensitivity with large arrays of SiPMs},
2020 JINST 15 P01008, doi: 10.1088/1748-0221/15/01/P01008.

\bibitem{DarkSideAmp1}
A. Razeto, et al.,
\emph{Very large SiPM arrays with aggregated output},
2022 JINST 17 P05038, doi: 10.1088/1748-0221/17/05/P05038.

\bibitem{DarkSideAmp2}
M. D'Incecco, et al.,
\emph{Development of a Very Low-Noise Cryogenic Preamplifier for Large-Area SiPM Devices},
IEEE TNS 65 4 2018, pp.1005-1011, doi: 10.1109/TNS.2018.2799325.


\bibitem{Cressler}
J.D. Cressler,
\emph{On the Potential of SiGe HBTs for Extreme Environment Electronics},
Proc. IEEE 93 1559, doi: 10.1109/JPROC.2005.852225.

\bibitem{RadekaAndCo}
S. Li, et al.,
\emph{LAr TPC Electronics CMOS Lifetime at 300 K and 77 K and Reliability Under Thermal Cycling}
IEEE Trans. Nucl. Sci., vol. 60, no. 6, pp. 4737-4743,
doi: 10.1109/TNS.2013.2287156.








\end{thebibliography}
\end{document}